\documentclass[genTeX]{nrc1}   
\usepackage[pctex32]{graphicx}

\volyear{??}{????}
\received{???}
\accepted{???}

\newcommand{\beq}{\begin{equation}}
\newcommand{\eeq}{\end{equation}}
\newcommand{\bea}{\begin{eqnarray}}
\newcommand{\eea}{\end{eqnarray}}
\newcommand{\eq}[1]{(\ref{#1})}

\begin{document}

\title{Hadronic effects in leptonic systems:
muonium hyperfine structure and
anomalous magnetic moment of muon}
\author{S. I. Eidelman}
\address{Budker Institute for Nuclear Physics and Novosibirsk University,
Novosibirsk, 630090, Russia}
\author{S. G. Karshenboim}
\address{D. I. Mendeleev Institute for Metrology, St. Petersburg, 189620, Russia and 
Max-Planck-Institut f\"ur Quantenoptik, Garching, 85748, Germany}
\author{V. A. Shelyuto}
\address{D. I. Mendeleev Institute for Metrology, St. Petersburg, 189620, Russia}


\shortauthor{Eidelman et al.}

\maketitle

\begin{abstract}
Contributions of hadronic effects to the muonium physics and anomalous magnetic 
moment of muon are considered. Special attention is paid to 
higher-order effects and the uncertainty related to the hadronic 
contribution to the hyperfine structure interval in the ground 
state of muonium.
\end{abstract}
\begin{resume}
French version of abstract (supplied by CJP)
   \traduit
\end{resume}

\section{Introduction}

Quantum electrodynamics (QED) is a theory which covers all
interactions of leptons (electrons and muons) with photons. 
However, it may be not sufficient even for pure
leptonic systems that are not free of hadronic effects appearing
because of hadronic intermediate states. 
Such effects cannot be
calculated {\em ab initio} and additional
data on these states are needed. Since the data 
can be achieved mainly from experiment, their availability  and 
uncertainty impose some principal limits on any  QED calculations. 

Recently a number of projects on intensive
muon sources have appeared \cite{sources} and 
a new generation of precision experiments on muonium may appear in 
near future. Therefore it is timely to investigate the  
principal limits of QED tests related to hadronic physics. Here we discuss 
hadronic effects in muonium and free muon.

\section{Hadronic effects in muonium hyperfine structure: leading 
contribution}

\begin{figure}
\caption{The leading hadronic contribution to the muonium hyperfine structure}\label{fig2}
\begin{center}
\includegraphics{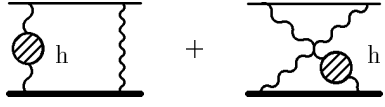}
\end{center}
\end{figure}

The leading hadronic contribution to the  
hyperfine splitting of the muonium  ground state 
comes from the diagram depicted in Fig.~\ref{fig2}.
We separate QED and hadronic effects in this diagram \cite{plb}
\beq \label{hadmu}
\Delta E_{\rm hadr}({\rm leading}) = -2\frac{\alpha^2}{\pi^2}
\frac{m_e m_\mu}{m_\mu^2-m_e^2}\frac{E_F}{1+a_\mu}
\int{ds}K_{\rm Mu}(s)\rho(s)
\eeq
and evaluate them independently. Here $K_{\rm Mu}(s)$ is a result of
the calculation of a two-photon
exchange diagram
with one massive photon ($s=\lambda^2$), the Fermi energy
\beq\label{efermi}
E_F={8\over 3}(Z\alpha)^4{m_e^2 m_\mu^2\over (m_\mu+m_e)^3} (1+a_\mu)
\eeq
is a nonrelativistic
value for the 
hyperfine structure interval, 
$s$ is the center-of-mass energy squared and
$\rho(s)$ is the hadronic spectral function:
\beq\label{Ree}
\rho(s)=\frac{2}{3s}R(s) =
\frac{\sigma(e^+e^-\to \gamma\to {\rm hadrons})}{2\pi\,\alpha^2}\;.
\eeq
The Fermi energy determined above contains the anomalous magnetic moment 
of muon, $a_\mu$, which is 
also affected by hadronic effects (see. Sect.~4).

\begin{figure}
\caption{The skeleton two-phonon exchange diagrams}\label{fig3}
\begin{center}
\includegraphics{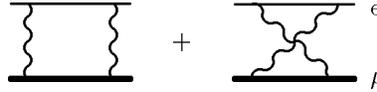}
\end{center}
\end{figure}

The basic expression has been obtained in Ref.~\cite{plb}. Here we briefly 
reproduce the derivation.
We start from the evaluation of the two-photon skeleton contribution 
(Fig.~\ref{fig3})
\beq\label{sce}
\Delta E_{\rm skel} = E_F~\frac{Z\alpha}{\pi}\frac{m_\mu m_e}{m_\mu^2-m_e^2}\int_{0}^
{\infty}{\frac{dk^2}{k^2}
\Big(J(k,m_\mu) - J(k,m_e)\Big)}\;,
\eeq
where
\[
J(k,\mu)=2\left[\frac{1}{k}\sqrt{k^2+4\mu^2}-1\right]-\frac{1}{4\mu^2}
\left[k\sqrt{k^2+4\mu^2}-k^2-2\mu^2\right]\;.
\]
We introduce hadronic polarization effects with a dispersion substitute 
for the photon propagator
\beq
\frac{1}{k^2}\to\frac{\alpha}{\pi}\int{\frac{d s~\rho(s)}{k^2+s}}\;.
\eeq
and after $k$-integration we obtain
\bea
K_{\rm Mu}(s)&=&
\left(\frac{s}{4m_\mu^2}+2\right)
\sqrt{1-\frac{4m_\mu^2}{s}}
\ln{\frac{1+\sqrt{1-\frac{4m_\mu^2}{s}}}{1-\sqrt{1-\frac{4m_\mu^2}{s}}}}\nonumber\\
&~&
-\left( \frac{s}{4m_\mu^2} +\frac{3}{2}\right)
\ln{\frac{s}{m_\mu^2}}+\frac{1}{2}+...\;.
\eea

\begin{figure}
\caption{Asymptotic behaviour of hadronic kernel. $\Delta K_{\rm Mu}(s)
=K_{\rm Mu}(s)-K_{\rm Mu}^{(0)}(s)$.}\label{fig56}
\begin{center}
\includegraphics{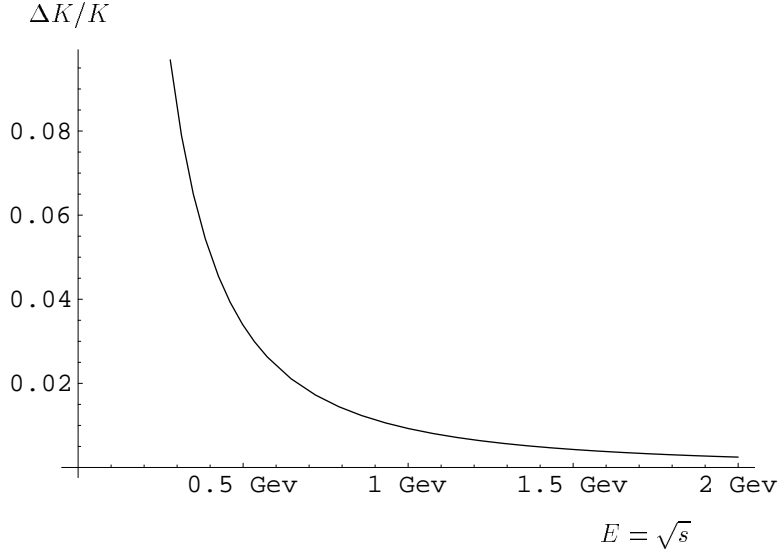}
\end{center}
\end{figure}

We are interested in relatively high $s$ and the expansion of the exact 
result
\bea\label{AsKmu}
K_{\rm Mu}(s)=
&-&\left\{\frac{4m_\mu^2}{s}\left[\frac{9}{8}\ln{\frac{s}{m_\mu^2}}
+\frac{15}{16}\right]
+\left(\frac{4m_\mu^2}{s}\right)^2\left[\frac{5}{16}\ln{\frac{s}{m_\mu^2}}
-\frac{17}{96}\right]\right.\nonumber\\
&~&
\left.+
\left(\frac{4m_\mu^2}{s}\right)^3
\left[\frac{21}{128}\ln{\frac{s}{m_\mu^2}}-\frac{73}{512}\right]
+\dots\right\}
\eea
is useful for further evaluations.
For $s=m_\rho^2$ the leading term
\beq
K_{\rm Mu}^{(0)}(s)=\frac{4m_\mu^2}{s}
\left[\frac{9}{8}\ln{\frac{s}{m_\mu^2}}+\frac{15}{16}\right]
\eeq
is a good enough approximation (within 2\%). 
For higher $s$ it is even better (see Fig.~\ref{fig56}).

A simple model  can be used to estimate  the leading hadronic contribution 
(cf. Ref.~\cite{sty}). The spectral function consists of
several narrow-pole contributions ($\rho$, $\omega$, $\phi$)
\beq\label{simple}
\rho_{\rm pole}(s)
= \sum_{\rm res}\frac{4\pi^2}{f_{\rm res}^2}\delta(s-M_{res}^2)
= \frac{3\pi\Gamma({\rm res}\to e^+e^-)}
{\alpha^2\;M_{\rm res}}\delta(s-M_{res}^2)
\eeq
and background with $R=2$ for 1 Gev$^2<s<$ 4 Gev$^2$ and $R=4$ for 
4 Gev$^2<s$. Here $M_{\rm res}, \Gamma({\rm res}\to e^+e^-)$ and
$f_{\rm res}$ are the resonance mass, leptonic width and coupling
constant, respectively.  
The results are summarized in Table~\ref{tab1}.
We estimate a systematic uncertainty of the model as 10\%. That is
not good enough for a calculation
of the leading term but sufficient for higher-order hadronic
corrections.

\begin{table}
\caption{Estimation of the leading hadronic contribution to the 
hyperfine structure in a simplified model.}\label{tab1}
\begin{center}
\begin{tabular}{ccc}
\hline*
Contribution &$\Delta E$ & $ \Delta E/\Delta E_{\rm hadr}$ \\
 &[kHz] & [\%] \\
\hline
$\rho$&0.151(7)&65(3)\\
$\omega$&0.013&5.5(2)\\
$\phi$&0.014&6.0(2)\\
$s=1-4$ Gev$^2$&0.045&19\\
$s > 4$ Gev$^2$& 0.009 & 3.8\\
\hline
Total&0.232(7)&99(3)\\
\hline*
\end{tabular}
\end{center}
\end{table}

Any realistic calculation of the leading hadronic term targeting a 
higher accuracy should deal 
with experimental data on e$^+$e$^-$ annihilation into hadrons 
(see e.g. Refs.~\cite{faustov,prd,narison}).
The most accurate recent result 
\cite{prd}\footnote{The result of Ref.~\cite{narison} has been obtained  
using $\tau$ lepton data and the accuracy quoted there is overestimated. 
The discussion of the possibility to use $\tau$ data can be found in 
Sect.~\ref{g-2}.} is 
\beq
\Delta E_{\rm hadr}({\rm leading}) = 0.233(3)~\mbox{kHz}\;.
\eeq

Since the publication of Ref.~\cite{prd}, some new experimental data appeared 
and the result can be now updated. These new data are:
\begin{itemize}
\item
The CMD-2 detector has recently published the final results
of their analysis for the reaction $e^+e^- \to \pi^+\pi^-$
in the c.~m.~energy range 610 to 960 MeV \cite{data1}.
Because of the new, more precise approach to calculating
radiative corrections which also include
a correction for the final state radiation, the
values of the cross section are slightly smaller than before.
\item
The new data on the $\phi$
meson leptonic width obtained by the CMD-2 and SND groups in 
Novosibirsk \cite{data2}.
\end{itemize}
As a result, the updated value of the leading order hadronic
contribution becomes 231.2 $\pm$ 2.9 Hz compared to our last
year value of 233.3 $\pm$ 3.1 Hz quoted above. We expect that after the 
analysis of other hadronic modes is completed in Novosibirsk, the precision
of the leading order contribution can be improved to about 2 Hz.

\section{Hadronic effects in muonium hyperfine structure: next-to-leading 
term}

\begin{figure}
\caption{The next-to-leading hadronic contribution to the hyperfine splitting in muonium}\label{fig5}
\begin{center}
\includegraphics{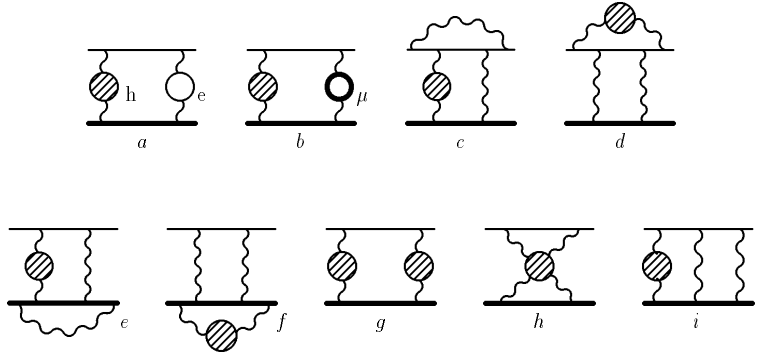}
\end{center}
\end{figure}

With an accuracy at a one-percent level for the leading hadronic 
contribution one has to take into account higher-order hadronic effects. 
The corresponding diagrams for some higher-order hadronic terms are 
presented in Fig.~\ref{fig5}. The first evaluation presented 
in Ref.~\cite{plb} gave a result
\beq
\Delta E_{\rm hadr}({\rm non-leading}) = 0.007(2)~\mbox{kHz}\;.
\eeq

Most of the contributions can be expressed in terms of corrections
to $K_{\rm Mu}$ in the form
\bea\label{klog}
\Delta K&=&
-{\alpha\over \pi}\frac{4m_\mu^2}{s}
\biggl\{\ln{\frac{9}{8}\frac{m_\mu^2}{m_e^2}}\ln{\frac{s}{m_\mu^2}}
+\frac{9}{8}\ln^2{\frac{s}{m_\mu^2}}
+\frac{15}{16}\ln{\frac{m_\mu^2}{m_e^2}}
+C_1\ln{\frac{s}{m_\mu^2}}+C_2
\biggr\}\nonumber\\
&\simeq&- {\alpha\over \pi}\frac{4m_\mu^2}{s}
\left\{47.6+17.7+10.0+C_1\cdot 4.0+C_2\right\}\;.
\eea
The numerical values are given for $s=m_\rho^2$. We present here
the calculation of double-logarithmic terms
and that of a single-logarithmic term with the bigger logarithm 
$\ln(m_\mu^2/m_e^2) \simeq 11 >
\ln(m^2_\rho/m_\mu^2) \simeq 4$. A calculation of other logarithmic
contributions is in progress.

The biggest contributions come from Fig.~\ref{fig5}{\em a}. The result is
\beq\label{ka}
\Delta K_{\rm a} = - 3{\alpha\over 3\pi}\frac{4m_\mu^2}{s}
\left\{
\left[\frac{9}{16}\ln^2{\frac{s}{m_\mu^2}}
-\frac{3\pi^2}{8} -\frac{27}{32} \right]
\right.
\left.
+\left[\frac{9}{8}\ln{\frac{s}{m_\mu^2}}+\frac{15}{16}\right]
\left[\ln{\frac{m_\mu^2}{m^2}}-\frac{5}{3}\right]
\right\}\;.
\eeq

Our final estimate is
\beq
\Delta E_{\rm hadr}({\rm non-leading})=0.005(2)\; {\rm kHz}\;.
\eeq
The contributions are summarized in Table~\ref{tab2}. The result is 
somewhat lower than in Ref.~\cite{plb} mainly because of single-logarithmic 
contributions from Fig.~\ref{fig5}{\em c} and {\em e}.

\begin{table}
\caption{Higher-order hadronic contributions to muonium hyperfine 
splitting. The results are presented in units of $\alpha/\pi \cdot E_{\rm hadr}({\rm leading})$.}\label{tab2}
\begin{center}
\begin{tabular}{cc}
\hline*
Correction & Contribution \\
& [$\alpha/\pi \cdot E_{\rm hadr}$]  \\
\hline$\Delta E_{\rm a}$ & 9.8 \\
$\Delta E_{\rm b}$ & 1.6 (7) \\
$\Delta E_{\rm c}$ &-1.0(3) \\
$\Delta E_{\rm d}$ & $\pm 0.1$ \\
$\Delta E_{\rm e}$ &-1.0(3)\\
$\Delta E_{\rm f}$ &$\pm 0.1$\\
$\Delta E_{\rm g}$ & 0.10(2)\\
$\Delta E_{\rm h}$ &$\pm 0.1$\\
$\Delta E_{\rm i}$ &$\pm 0.3$\\
\hline
$\Delta E_{\rm high}$ &  9.2(14)\\
\hline*
\end{tabular}
\end{center}
\end{table}

Recently higher-order hadronic effects were also studied in 
Ref.~\cite{faustov1}. In particular, the hadronic light-by-light 
contributions were under examination. The contribution of the diagram 
in Fig.~4{\em h} is 1.5~mHz and so it 
is consistent with our conservative estimation. A bigger contribution 
of 3.65~Hz found there (related 
to Fig. 2 of Ref. ~\cite{faustov1}) is actually due to the 
correction to the anomalous magnetic moment of muon and affects hyperfine 
splitting accordingly to Eq.~\eq{efermi}. Therefore it should be excluded from 
our consideration.

More accurate calculations are in progress and we hope to reduce
the uncertainty by a factor of 4 \cite{ksnew}.

\section{\label{g-2} Anomalous magnetic moment of muon}

\begin{figure}
\caption{Leading hadronic contribution to the anomalous magnetic 
moment of muon}\label{fig1}
\begin{center}
\includegraphics{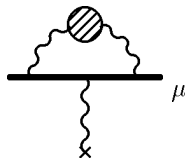}
\end{center}
\end{figure}

Since the 2001 publication of the E821 result on the
anomalous magnetic moment of muon \cite{amu},
a lot of efforts have been put on the new estimation of the
hadronic correction to the muon anomaly, ${{a}}^{had}_{\mu}$,
see e.g. Ref.~\cite{ynd}. However, these works were based on 
either not final experimental data or involved various theoretical
assumptions. In view of the extreme importance of the E821 result and
possible indications to deviation from the Standard Model
it is necessary to have a model-independent analysis based exclusively
on the data, similar to that of Ref.~\cite{ej}. 
The new calculation based on the most recent data
from $e^+e^-$ annihilation at low energy in Novosibirsk and Beijing as
well as from $\tau$ decay data, which also includes
new developments in theory, has recently been published \cite{enew}. 
It showed a real 
breakthrough in the accuracy of the    
${{a}}^{had}_{\mu}$ estimation which became possible
after the final data on $e^+e^- \to \pi^+\pi^-$ from CMD-2
with a 0.6\% systematic uncertainty have been published \cite{data1} and, 
in addition, the final analysis of the R measurement by BES between 2 and
5 GeV appeared \cite {data3}. The resulting accuracy of the estimate
increases dramatically and reaches $\simeq$ $70\cdot10^{-11}$ based
on $e^+e-$ data only. This accuracy is much better compared to the
uncertainty of $153\cdot10^{-11}$ obtained in 1995 in 
Ref.~\cite{ej} and is very close to that obtained in the 1998 analysis 
of Ref.~\cite{dh} which 
also included the $\tau$ data and QCD sum rules. Calculations show that 
another significant improvement (by a factor of about 1.5) is
possible when high precision $\tau$ lepton decay data are also used. 
However, for their reliable use one should first understand the 
reason of the observed discrepancy between two data sets: 
the two pion spectral functions obtained from $e^+e^-$ and $\tau$ 
lepton data do not agree confirming earlier evidence for this effect 
~\cite{two}. 
The solution of this problem should also 
involve a thorough investigation of the effects of isospin breaking 
corrections as well as additional radiative effects in $\tau$ 
decays \cite{tau}.

The real uncertainty of the theoretical estimate will also strongly
depend on the understanding of the role of higher-order effects.
These include diagrams with two loops which can be reliably
calculated from the $e^+e^-$ data using dispersion relations \cite{kra}
and much more complicated effects due to hadronic light-by-light scattering. 
At the present high level of accuracy, both effects give a quite
substantial contribution of the order of $100\cdot 10^{-11}$ each, i.e. 
about or even larger than the current uncertainty of the 
leading order hadronic 
contribution. Recent reevaluations of the pseudoscalar pole contribution
dominating in the hadronic light-by-light scattering diagrams revealed that 
its sign was previously wrong \cite{lbl1}. This initiates new 
calculations of the higher-order effects and more realistic estimates of 
their accuracy \cite{lbl2}.

\section{Summary}

The hadronic effects become an unavoidable part of theory of ``pure'' 
QED quantities, such as the anomalous magnetic moment of muon and 
hyperfine interval in the ground 
state of muonium. In the case of muon g-2 they limit present accuracy of the theoretical prediction whereas for muonium their impact is more relevant for the future experiments with intensive muon sources and thus higher expected accuracy.

The new experimental result on the anomalous magnetic moment of muon has
recently appeared \cite{freshmu}. It has about the same accuracy as
theory values from $e^+e^-$ and/or $\tau$ data. A more accurate 
result is expected soon. The hadronic contributions 
dominate in the theoretical uncertainty and require a more detailed study. 
On the contrary, no results on muonium hyperfine 
structure are expected in near future and theory is more accurate 
than the experiment. However, 
to clearly understand what level of accuracy could  be a target of new 
experiments on muonium, one should study theoretical uncertainties. 
The bound state QED theory is overviewed in Ref.~\cite{prd} and some 
progress for {\em ab initio} calculations is possible. Better 
understanding of hadronic effects 
is needed to find a level of accuracy which cannot be superseded by 
the {\em ab initio} QED calculation. Several projects on the calculation
of hadronic effects in physics of muon and muonium which are 
shortly described here are in progress and we hope to report on 
their results  soon.

\section*{Acknowledgements}
The authors would like to thank Andrzej Czarnecki, Michel Davier,
Andreas H\''{o}cker, Fred Jegerlehner, Klaus Jungmann and Andrea Vacchi
for stimulating discussions. This work
was supported in part by the RFBR grant 00-02-16718.

\end{document}